\documentclass[11pt]{article}

\usepackage[T1]{fontenc}
\usepackage{lmodern}
\usepackage{microtype}
\usepackage[utf8]{inputenc}
\usepackage{authblk}
\usepackage{amsmath}
\usepackage{graphicx}
\usepackage{listings}
\usepackage{xcolor}
\usepackage[margin=1in]{geometry}
\usepackage[colorlinks=true,citecolor=blue,linkcolor=blue,urlcolor=blue]{hyperref}

\title{Tamper-Proofing with Self-Modifying Code}
\author[1]{Gregory Morse\\\texttt{morse@inf.elte.hu}}
\author[1]{Tam\'{a}s Kozsik\\\texttt{kto@elte.hu}}
\affil[1]{E\"{o}tv\"{o}s Lor\'{a}nd Tudom\'{a}nyegyetem / University (ELTE), Budapest, Hungary}
\date{}

\begin{document}
\pagestyle{empty}
\maketitle
\thispagestyle{empty}

\begin{abstract}
Classical computability theory tells us that self-modifying code (SMC) on a
deterministic universal Turing machine can be simulated by non-SMC code on the
same model.  That abstraction, however, omits the external timing inputs,
concurrency, and microarchitectural state that dominate practical execution on
modern processors.  We argue that once timing, ordering, and self-introspective
effects are treated as observables, a \emph{practically faithful} non-SMC
reproduction of timed SMC becomes detectably expensive on commodity systems.
We present a tamper-proofing model that combines introspective and polymorphic
SMC, reliable clocks, and runtime timing predicates to bind integrity checks to
execution behavior.  We distinguish static and dynamic SMC generation,
characterize the timing semantics needed to avoid catastrophic pipeline clears,
and give x86-64 design primitives for checksum-driven self-patching.  We also
report timer measurements, performance comparisons, and performance-monitoring
counter evidence showing that careful engineering---especially loop unrolling and
cross-page modification---substantially reduces the overhead of SMC while
preserving its tamper-detection value.  The paper concludes with an efficiency
analysis, a threat model, and deployment guidance for trusted code executing in
untrusted environments.
The threat model is formalised in Lean~4, relative to an explicit platform contract: the
security theorem, the necessity of each assumption and which assumptions each adversary level
satisfies or breaks are machine-checked, and the development is registered with the Palomar registry.
\end{abstract}


\section{Introduction}\label{sec:introduction}

It is classical in computability theory that self-modifying code (SMC) on a
deterministic universal Turing machine (UTM) can be simulated by non-SMC code
on the same model; SMC does not increase computational power beyond Turing
completeness, but rather changes representation and potentially efficiency
\cite{10.5555/1095587,hopcroft1979introduction,Arora_Barak_2009}.  That
baseline, however, abstracts away \emph{external inputs}.  In practice, time
sources such as Coordinated Universal Time (UTC) and processor time-stamp
counters (TSCs), together with concurrent scheduling, introduce nondeterminism:
even if a TSC were made deterministic, UTC would not be; in parallel settings,
TSC readings are further complicated by cache and memory synchronization.

Once such inputs are admitted as observables, the equivalence of SMC to
non-SMC becomes fragile in a tamper-proofing context.  Functional behavior can
still be simulated in principle, but preserving \emph{timing}, \emph{ordering},
and \emph{self-introspective effects} in the presence of an adversary generally
cannot be achieved without either reintroducing self-modification or incurring
detectable distortions.

\paragraph{External inputs: what helps, what does not.}
Many machine and system identifiers---CPU ID strings, NIC MAC addresses,
disk/RAM configurations, or operating-system fingerprints---are useful for
binding a license to a host \cite{tan2013}, but they solve fingerprinting more
directly than tamper-proofing and are operationally brittle.  System-activity
signals such as mouse movement or event logs either feed entropy pools
\cite{HU20092286} or exhibit regularity that attackers can mimic.  Cloud- or
server-provided data can be incorporated as an external input, but replay and
protocol attacks complicate correctness unless additional machinery is deployed;
moreover, such channels do not protect code at the machine-architecture level
against manual debugging, thread injection, or remote process-memory writes.  The mechanism proposed here does not prevent these either: it detects persistent modification of the checksummed bytes and interference inside timed intervals, but not injected code that leaves the checksummed bytes intact, register tampering outside timed intervals, or a modification that is undone before the next covering check (Section~\ref{sec:assumptions}).  A
well-engineered network time source can, however, be a reliable UTC input.

\paragraph{Practical stance.}
For integrity checks that bind behavior to time, there is effectively no
\emph{practically faithful} non-SMC equivalent that preserves both semantics and
timing in the face of an adversary on contemporary systems.  An attacker who controls the time source need not model the processor at all: replaying readings recorded from native runs makes any threshold predicate on those readings pass.  Timing-based detection therefore requires a time source outside the adversary's control, for example \texttt{RDTSC}/\texttt{RDTSCP} read inside the protected code against a user-mode adversary, or an external source against a kernel-level adversary.  An adversary who must run the protected code under emulation with an honest clock needs a deterministic, fully characterizable timing model, and modern commodity processors do not admit one.
Execution timing is influenced by partially undocumented microarchitectural
state---branch predictors, caches, coherence protocols, speculative execution,
microcode---together with concurrent activity and operating-system scheduling.
Reproducing these effects within tight timing tolerances would require a faithful
simulation of both the processor and the operating system, including their
dynamic state, and doing so at practical speed is infeasible.

\paragraph{Model.}
We propose a tamper-proofing model in which SMC is (i) \emph{introspective}, in
that it checksums its own code, and (ii) \emph{polymorphic}, in that it rewrites
itself while executing.  The model uses algorithmic kernels whose control flow and instruction count are independent of the data to
reduce unrelated timing leakage and may treat sensitive data as code
(SMC-backed storage) to harden against data-only faults.  A virtualized SMC
wrapper can dynamically generate an inner SMC layer.  Formally, we assume a
timer $T(t)$ chosen per query $t$ for reliability and a checksum function $C(c)$
over code or data $c$; the predicate $P(T,C)$ returns true for the unmodified protected code (when its timing lies within the calibrated bounds) and false for every modification that changes the checksum state or violates the time bound.

\paragraph{Relation to prior work.}
Tamper-proofing via watermarking is well studied \cite{1027797}.  Self-check-
summing with SMC has also been explored \cite{1565232}, but typically with
static checksum routines rather than checksum kernels that rewrite themselves
while executing.  Remote runtime-integrity schemes can rely on a verifier in
the cloud \cite{8795389}; our goal is to avoid such infrastructure and keep the
assurance mechanism local to the untrusted host.  This inverts the conventional
``trustworthy computing'' perspective, which usually assumes a trusted host
running untrusted software.  Our setting instead asks how trusted software can
retain integrity on an untrusted host, which is the operationally harder
direction.

\paragraph{Contributions.}
The paper makes four concrete contributions.
\begin{itemize}
  \item It frames timed, introspective SMC as a practical tamper-proofing
  primitive whose non-SMC reproduction is detectably costly on modern systems.
  \item It develops a structured engineering model covering reliable clocks,
  timing semantics, static versus dynamic generation, and concrete x86-64 design
  primitives.
  \item It reports empirical timing and performance-counter results that explain
  when SMC is impractical and when it becomes efficient enough to deploy.
  \item It states explicit efficiency and security assumptions so that the scope
  and limitations of the approach are clear.
\end{itemize}

\begin{figure}[t]
  \centering
  \includegraphics[width=0.72\linewidth]{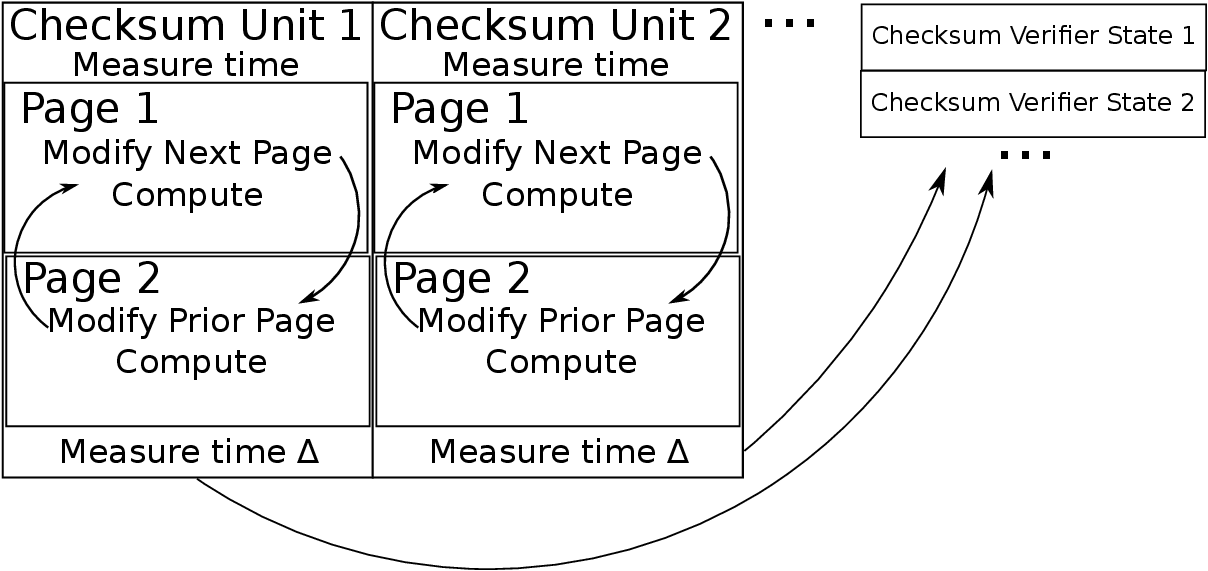}
  \caption{Design overview of checksum units and state verification.}
  \label{fig:tamperproof}
\end{figure}

Figure \ref{fig:tamperproof} gives a high-level overview of the resulting design.

The remainder of the paper proceeds as follows.  Section \ref{sec:clocks}
evaluates practical time sources.  Section \ref{sec:static-dynamic} contrasts
static and dynamic SMC generation.  Section \ref{sec:timing-semantics}
formalizes the timing semantics relevant to detection.  Sections
\ref{sec:static-primitives} and \ref{sec:dynamic-primitives} present static and
dynamic design primitives.  Section \ref{sec:performance} reports measured
performance.  Section \ref{sec:efficiency} interprets the cost model.  Section
\ref{sec:using-primitives} shows how the primitives compose into a deployment
scheme.  Section \ref{sec:assumptions} states the threat model and assumptions,
and Section \ref{sec:summary} concludes.

\section{Reliable clocks}\label{sec:clocks}

The read TSC instructions \texttt{RDTSC} and \texttt{RDTSCP} are not perfectly
reliable in isolation because modern processors expose hyper-threading,
out-of-order execution, speculative execution, and cache effects that can
perturb measurements \cite{6557172}.  Nevertheless, they can be made reliable
enough statistically by repeated execution and by surrounding them with
serialization instructions such as \texttt{CPUID} and \texttt{RDTSCP}.  In many
tamper-proofing contexts, a high-probability integrity decision is sufficient to
continue execution.

Most contemporary systems also expose several other clocks.  High Precision Event
Timer (HPET), real-time clock (RTC) facilities, and operating-system monotonic
counters provide alternative timing sources with different resolution and noise
profiles.  The specific x86 instructions discussed here are not unique in spirit:
AVR32, SPARC V9, PowerPC, ARMv7, and ARMv8 all provide related counter-style
instructions, underscoring the near universality of hardware time measurement.

For x86 time-stamp counters, the instructions \texttt{RDTSC} and \texttt{RDTSCP}
have the corresponding intrinsics \lstinline|__rdtsc| and \lstinline|__rdtscp|,
which are readily usable from user mode.  On many systems they offer the most
fine-grained timing information available to ordinary software.

On Windows, several Application Programming Interface (API) calls expose timer
state.  UTC can be queried with
\lstinline|GetSystemTimePreciseAsFileTime|.  The high-resolution performance
counter is obtained with \lstinline|QueryPerformanceCounter| and calibrated with
\lstinline|QueryPerformanceFrequency|.  The interrupt timer is available through
\lstinline|QueryInterruptTimePrecise|.  These three are practically in the same
precision class.  Coarser timers include the tick counter
\lstinline|GetTickCount64()| and multimedia timers via
\lstinline|timeBeginPeriod(1)|, \lstinline|timeEndPeriod(1)|, and
\lstinline|timeGetTime()|, though they are not intended for precise
microbenchmarking.

On Linux, \lstinline|clock_gettime| provides access to several clocks.  The
system clock is obtained through \lstinline|CLOCK_REALTIME|.  A monotonic clock
analogous to an interrupt timer is measured with
\lstinline|CLOCK_MONOTONIC|, and a boot-time clock including suspended time is
available through \lstinline|CLOCK_BOOTTIME|.  UTC time can also be queried with
\lstinline|gettimeofday|, although \lstinline|clock_gettime| is generally the
cleaner interface in contemporary software.  Standard C and C++ library
abstractions wrap part of this functionality and may be easier to use when the
implementation details do not matter.

\section{Static versus dynamic generation of SMC}\label{sec:static-dynamic}

Static SMC is handwritten assembly code linked into the program at compile time.
It is difficult to write and maintain because assemblers and C/C++ toolchains do
not directly accommodate code that rewrites its own instruction bytes.  Compiler
support is also inconsistent: Microsoft Visual C++ supports inline assembly on
x86 but not x86-64, whereas the GNU Compiler Collection allows inline assembly
more broadly.

Dynamic SMC is generated at runtime, for example with an assembler library such
as AsmJit \cite{asmjitAsmJit}.  Although such frameworks are typically used for
just-in-time compilation, they also provide a convenient foundation for SMC that
must be polymorphic or context sensitive.  Features missing from conventional
assemblers can be introduced programmatically, and the generated code can adapt
to the operating-system version, processor family, address-space layout, or
deployment policy at runtime.

Static code is normally loaded into read-execute (RX) pages.  Since SMC needs
write access, the binary or loader configuration must be adjusted accordingly.
Under MSVC, this can be expressed at link time with
\lstinline|/SECTION:.text,RWE|.  With other toolchains, such as GCC on Portable
Executable (PE) or Executable and Linkable Format (ELF) targets, the required
section permissions may need manual editing in the binary format metadata.

For dynamically generated SMC, the memory allocator must request pages with the
appropriate permissions.  On Windows, this is achieved with
\lstinline!VirtualAlloc(NULL, codeSize, MEM_RESERVE | MEM_COMMIT, PAGE_EXECUTE_READWRITE)!.
On Linux, the corresponding mechanism is typically \lstinline|mmap|.  In both
cases, deployment policies should be reviewed carefully if write-execute (WX/RWX)
pages are restricted.

Dynamic SMC also makes address tracking more difficult for emulation tools.
Because the code layout is generated at runtime, the addresses that a semantics-
faithful non-SMC emulator must monitor are themselves contextual.  This is one of
the reasons dynamic generation is especially attractive for tamper-proofing.

\section{Timing semantics of SMC}\label{sec:timing-semantics}

Efficient tamper-proofing with SMC requires careful attention to modern
microarchitectural behavior, particularly instruction pipelines and memory
hierarchies.  Most processor optimizations of recent decades have focused on deep
pipelines, speculative pre-processing, and aggressive caching.  Na"ive SMC can
invalidate that work and degrade execution to something resembling a much older
processor running at a higher clock frequency.  For a tamper-proofing scheme,
such a penalty is unacceptable.

A practical workaround is to structure the checksum routine so that it does not
modify the currently executing instruction stream.  \emph{Loop unrolling} is one
useful technique: it increases the number of instructions executed between
modification points and thus allows the pipeline to remain populated.  When cache
effects matter, unrolling can also distribute self-modifying regions across
distinct pages, reducing the chance that the executing page is also the one being
written.

Although modern processors ultimately remain von Neumann machines, their deep
pre-execution machinery makes SMC expensive when it touches instructions that are
already in flight.  On x86, such cases can trigger a \emph{machine clear}: the
processor squashes in-flight instructions and restarts fetch/decode for the
logical processor.  This is a major source of slowdown.  Similar effects can also
appear in cross-modifying code (XMC) settings where one thread writes code later
executed by another.  Designs therefore have to account for these penalties and
avoid them when possible.

The detection of SMC by the processor is itself approximate.  It often uses the
current page or a smaller region as the detection unit, which means false
positives can occur when a modification is near, but not identical to, the bytes
currently in flight.  Consequently, both timing thresholds and code layout must
be engineered with these approximations in mind.

Hardware performance counters help tune these designs.  On x86, model-specific
registers (MSRs) can configure event counters that are later read through
\texttt{RDPMC} if control register 4 allows it.  The configuration interface is
vendor specific in its addresses but conceptually similar across Intel and AMD
processors \cite{intel2016,amd2023}.  Table \ref{tbl:smcx8664} summarizes the key
register and event identifiers used in our measurements.

Between Windows and Linux, the main difference is access control.  Performance
counter configuration is privileged.  On Windows, practical experimentation may
therefore require a signed kernel driver such as WinRing0 \cite{openlibsysWhatapossWinRing0}.
On Linux, \lstinline|perf_event_open| can expose performance counters to user mode
under policy, so a separate kernel module is not always necessary.  Importantly,
the counters were used only for \emph{measurement and tuning}; deployment of the
tamper-proofing mechanism itself does not require direct PMC access.

\begin{table}[t]
\centering
\small
\begin{tabular}{|c|c|p{4.3cm}|}
\hline
\textbf{Item} & \textbf{Intel x86-64} & \textbf{AMD64} \\ \hline
Perf. Counter Global Control & 0x38F & 0xC0000301 \\ \hline
Perf. Event Selector 0 & 0x186 & 0xC0010000 \\ \hline
Perf. Counter 0 & 0xC1 & 0xC0010004 \\ \hline
SMC Event Name / Unit Mask & \texttt{MACHINE\_CLEARS.SMC} & \parbox[t]{4.3cm}{\raggedright\scriptsize\texttt{PIPELINE\_RESTART\_DUE\_TO\_\\SELF\_MODIFYING\_CODE}} \\ \hline
SMC Event Value & 0xC3 / 0x4 & 0x21 / 0x0 \\ \hline
\end{tabular}
\caption{Intel x86-64 and AMD64 address and event reference values for representative supported microarchitectures.}
\label{tbl:smcx8664}
\end{table}

\section{Static SMC tamper-proofing design primitives}\label{sec:static-primitives}

To demonstrate the core idea, we begin with a simple static design.  Efficient
SMC depends heavily on the details of the instruction set architecture (ISA).
Many ISAs contain opcode families whose encodings differ only in a few bits.
On x86-64, this makes it possible to exchange arithmetic operations by toggling a
small number of bits in place.  In particular, several long-standing arithmetic
instructions share a common format for a 64-bit register destination and a 64-bit
register-or-memory source.  Table \ref{tbl:instgroup} shows the useful subset.

By exploiting this encoding pattern and combining it with loop unrolling, we
develop a checksum scheme.  Here, a \emph{checksum scheme} is a self-verification
mechanism in which a program computes a lightweight integrity value over a region
of its own code or data and uses that value at runtime to detect tampering or to
select the next execution state.  The checksum is not intended to be collision
resistant in the cryptographic sense; the protection comes from coupling it to
dynamic state and timing.

With multiple checksum functions, the program can maintain validation data for a
set of reachable states defined by each checksum kernel's last modification.
Reset points before or after initialization keep the state space manageable.
Figure \ref{fig:staticsmc} shows a minimal SMC checksum kernel, while Figure
\ref{fig:csmc} shows a C++ implementation that emulates the same behavior as
faithfully as possible without using SMC.

One x86-64 feature is especially helpful here: instruction-pointer-relative
addressing.  Because RIP-relative addressing allows nearby code bytes to be
reached without first materializing an absolute address in a register, the SMC
kernel can patch itself with comparatively little scaffolding.  Earlier 16-bit
and 32-bit x86 variants do not provide this mechanism in the same form; there,
the programmer typically needs an auxiliary register and a \texttt{CALL}/\texttt{POP}
style idiom to recover the current address, which perturbs the stack and makes
the checksum kernel less elegant.

\begin{table}[t]
\centering
\small
\begin{tabular}{|c|c|c|c|c|}
\hline
\textbf{Instruction} & \textbf{Opcode} & \textbf{Opcode[3:5]} & \textbf{LAT} & \textbf{RCP} \\ \hline
ADC & 0x13 & 2 & 1 & 1 \\ \hline
ADD & 0x03 & 0 & 1 & 0.25 \\ \hline
AND & 0x23 & 4 & 1 & 0.25 \\ \hline
CMP & 0x3B & 7 & 1 & 0.2 / 0.25 \\ \hline
OR  & 0x0B & 1 & 1 & 0.25 \\ \hline
SBB & 0x1B & 3 & 1 & 1 \\ \hline
SUB & 0x2B & 5 & 1 & 0.25 \\ \hline
XOR & 0x33 & 6 & 1 & 0.25 \\ \hline
\end{tabular}
\caption{x86 arithmetic instruction pattern used by the checksum kernel.  Latency and reciprocal throughput are in cycles; ranges reflect microarchitectural variation \cite{asmjitAsmGrid}.}
\label{tbl:instgroup}
\end{table}

\begin{table}[t]
\centering
\small
\begin{tabular}{|c|c|}
\hline
ADC & Add with carry \\ \hline
SBB & Subtract with borrow \\ \hline
XOR & Exclusive OR \\ \hline
SHL & Shift left (logical, same as SAL) \\ \hline
LEA & Load effective address \\ \hline
SETcc & Set byte on condition \\ \hline
CMP & Compare operands \\ \hline
Jcc & Conditional jump \\ \hline
cc=S & Sign / negative (SF = 1) \\ \hline
cc=G & Greater than (signed, ZF = 0 and SF = OF) \\ \hline
cc=P & Parity (PF = 1) \\ \hline
\end{tabular}
\caption{Reference for instructions used by the example checksum kernel.  The parity flag uses the legacy low-8-bit parity computation.}
\label{tbl:instinfo}
\end{table}

\begin{figure}[t]
\centering
\begin{lstlisting}
checksum:
ADC RBX, QWORD PTR [RSI] ; 48 13 1E
SETS DL ; 0F 98 C2
SETP AL ; 0F 9A C0
SHL AL, 2 ; C0 E0 02
OR DL, AL ; 0A D0
SHL DL, 3 ; C0 E2 03
XOR BYTE PTR [$-2-3-3-3-2-3], DL
LEA RSI, [RSI+8]
CMP RDI, RSI
JG checksum
\end{lstlisting}
\caption{A simple static SMC checksum variant.}
\label{fig:staticsmc}
\end{figure}

\begin{figure}[t]
\centering
\begin{lstlisting}[language=C++]
const size_t offsetInFunc = 0xF;
// manually found based on MOV, LEA, XOR, CLC, ADC prefix
const PULONG64 pModOffset = (PULONG64)((ULONG64)((PBYTE)pTPCompOffset + offsetInFunc) & ~0x7ULL);
const unsigned char pModByteOffset = ((ULONG64)((PBYTE)pTPCompOffset + offsetInFunc) & 7) * 8 + 3;
// convert to bits, add 3 bits for the ADC/SBB/XOR/CMP offset
const unsigned __int64 preserve = ~(5ULL << pModByteOffset);
for (LPBYTE pEnd = (LPBYTE)pData + ulSize; (LPBYTE)pData < pEnd; pData++) {
    unsigned __int64 res;
    unsigned __int64 pDataVal = pData == pModOffset ?
        (*pData & preserve) |
        ((unsigned __int64)op << pModByteOffset) : *pData;
    switch (op) {
    case 0: res = sum += pDataVal; break;
    case 1: res = sum -= pDataVal; break;
    case 4: res = sum ^= pDataVal; break;
    case 5: res = sum - pDataVal; break;
    default: res = 0;
    }
    op ^= (!paritybyte((unsigned char)res) << 2) |
        ((res & 0x8000000000000000ULL) != 0);
}
\end{lstlisting}
\caption{Minimal C++ code to emulate the SMC checksum computation precisely.  The function arguments are the start address \lstinline|pData|, the protected length \lstinline|ulSize| (a multiple of 8 bytes) and the result address \lstinline|pResult|; \lstinline|pTPCompOffset| is the address of the body of the kernel \lstinline|TPComp| and is computed before the loop (not shown), \lstinline|sum| is stored to \lstinline|*pResult|, and \lstinline|op| is the persistent opcode state. The emulation is exact for an 8-byte-aligned region when \lstinline|op| equals the opcode state of the kernel at entry.}
\label{fig:csmc}
\end{figure}

The next static variant loop-unrolls the checksum body and modifies an opcode at
some fixed offset from each unrolled position.  Unrolling has two advantages.
First, it improves timing behavior: with a suitable offset and sufficient depth,
pipeline clears can be reduced dramatically and may disappear altogether.  Second,
it makes faithful non-SMC emulation harder because the emulator must track more
potential modification sites as the unroll depth grows.

\section{Dynamic SMC tamper-proofing design primitives}\label{sec:dynamic-primitives}

Dynamic generation generalizes the same ideas while removing many of the practical
limits of handwritten assembly.  Figure \ref{fig:dynsmc} shows a representative
AsmJit fragment.  Important features include the ability to bind labels at
arbitrary offsets, compute code sizes for fragments before final emission, and
emit RIP-relative x86-64 addressing directly.  These features are especially
valuable because x86-64 relative encodings can change size depending on the final
distance to a label, making precise layout difficult to calculate by hand.

\begin{figure}[t]
\centering
\begin{lstlisting}[language=C++]
if (lbl) cb.bind(*lbl);
cb.adc(x86::rbx, x86::ptr(x86::rcx));
cb.sets(x86::dl);
cb.setp(x86::al);
cb.shl(x86::al, 2);
cb.or_(x86::dl, x86::al);
cb.shl(x86::dl, 3);
cb.xor_(destlbl ? x86::byte_ptr(*destlbl, 1) :
    x86::byte_ptr(x86::rip, PAGE_SIZE), x86::dl);
// add 1 to skip the prefix byte
cb.lea(x86::rcx, x86::ptr(x86::rcx, 8));
\end{lstlisting}
\caption{Example snippet of dynamic SMC generated via AsmJit.}
\label{fig:dynsmc}
\end{figure}

Our implementation used three code blocks: a prologue, the checksum body, and an
epilogue.  The prologue and checksum body were first emitted with placeholder
addresses to determine exact code size.  Then the prologue and the required
number of checksum copies were arranged to fill two pages of memory, after which
the epilogue was appended and the whole region emitted into RWX pages.  The
prologue and epilogue preserve registers and satisfy calling conventions.

In practice, a dual-page layout with each unrolled checksum unit writing to the
corresponding offset in the alternate page proved sufficient and close to ideal.
We also used SMC to implement branch-free termination so that each unrolled
iteration avoids conditional control flow.  The final termination patch can still
induce an occasional pipeline clear, but this is infrequent enough to resemble a
branch misprediction rather than a persistent cost.

When executing the design, we logged processor and layout details to make the
experimental configuration explicit.  Figure \ref{fig:expsetup} shows a typical
output block.  Most of these values are discovered through \texttt{CPUID}, page
size queries, or the layout computations already described.

\begin{figure}[t]
\centering
\begin{lstlisting}
Manufacturer: GenuineIntel
DisplayModel: 0x9E DisplayFamily: 0x6
Architecture Performance Monitoring Version: 4 Number of Performance Counters per Logical Processor: 4 PMC bit width: 48
The page size for this system is 4096 bytes.
Code segment base: 0x7FF7B4521000 size: 0x8200 End address: 0x7FF7B4529200
Number of pages: 2 Unrolled loop size per page: 148 Intro bytes: 0x5F Code size: 0x1B
\end{lstlisting}
\caption{Representative processor, operating-system, and dynamic-layout details used during experimentation.}
\label{fig:expsetup}
\end{figure}

\section{Performance results}\label{sec:performance}

For the basic static SMC model, the measured execution time is on average
approximately $7.9\times$ slower than the non-SMC variant shown in Figure
\ref{fig:csmc}; the measurements are given in Table \ref{tbl:staticresult}.
Among the clocks tested, \texttt{RDTSCP} is the most informative and the most
stable for fine-grained timing.  Each timed interval also contained one console write and one performance-counter driver call, which add the same kind of overhead to every variant.  The minimum, average, and maximum values reveal
heavy-tailed noise caused by infrequent context switches, but the central mass of
the samples remains tight enough to support practical thresholding.

This distinction matters when interpreting the tables.  The maximum values should
not be read as the ``typical'' execution time of the checksum; they are dominated
by rare scheduler interference and other system-level disturbances.  The minima,
by contrast, are much more stable and better represent the uncontended baseline.
For deployment, this suggests calibrating thresholds from empirical quantiles or
repeated-sample windows rather than from raw maxima alone.

\begin{table}[t]
\centering
\small
\begin{tabular}{|c|c|c|c|c|c|c|}
\hline
Timer & \multicolumn{3}{c|}{SMC} & \multicolumn{3}{c|}{Non-SMC} \\ \hline
 & Avg. & Min. & Max. & Avg. & Min. & Max. \\ \hline
RDTSC & 8305286 & 7466370 & 17053914 & 1052981 & 959776 & 11039730 \\ \hline
RDTSCP & 8305389 & 7466412 & 17054428 & 1053077 & 959834 & 11040460 \\ \hline
Interrupt Time & 32044 & 28806 & 65800 & 4064 & 3704 & 42610 \\ \hline
Perf. Counter & 32045 & 28807 & 65802 & 4064 & 3703 & 42617 \\ \hline
Precise UTC & 32045 & 28806 & 65803 & 4065 & 3704 & 42622 \\ \hline
Tick Count & 3 & 0 & 16 & 0 & 0 & 16 \\ \hline
Multimedia & 3 & 1 & 7 & 0 & 0 & 4 \\ \hline
\end{tabular}
\caption{$10{,}000$ runs on a hexa-core Intel Core i7-9750H with 16~GB RAM under Windows~10; the checksum covered the entire 32.5~KiB (0x8200-byte) code segment.}
\label{tbl:staticresult}
\end{table}

The dynamically generated SMC variant is substantially more competitive.  As
shown in Table \ref{tbl:dynresult}, it is approximately $90.5\times$ faster than
the semantics-faithful, timing-constrained non-SMC variant and about $17.8\times$
faster than the basic static SMC implementation (Table~\ref{tbl:staticresult}).  This version used two pages and wrote
to the corresponding offset in the opposite page; using three or four pages did
not materially improve results.

The spread in the dynamic measurements also gives a rough sense of how often the
operating system perturbs the benchmark.  Using the \texttt{RDTSCP} figures as a
back-of-the-envelope estimate, the SMC variant yields
\[
\frac{10{,}000 \cdot (467{,}286 - 414{,}188)}{2{,}158{,}056 - 414{,}188} \approx 304,
\qquad
\frac{304}{10{,}000} \approx 3.0\%,
\]
which suggests that only a small fraction of runs suffer a substantial task-switch
style disturbance.  Performing the same estimate for the semantics-faithful
non-SMC variant gives roughly $7.8\%$ (Table~\ref{tbl:dynresult}, non-SMC columns, with the minimum subtracted in the same way).  The exact percentages should not be
over-interpreted, but they reinforce the practical point: the dominant source of
variance is scheduler noise, not instability in the SMC design itself.

\begin{table}[t]
\centering
\small
\begin{tabular}{|c|c|c|c|c|c|c|}
\hline
Timer & \multicolumn{3}{c|}{SMC} & \multicolumn{3}{c|}{Non-SMC} \\ \hline
 & Avg. & Min. & Max. & Avg. & Min. & Max. \\ \hline
RDTSC & 467241 & 414132 & 2157922 & 42300670 & 39599882 & 74294904 \\ \hline
RDTSCP & 467286 & 414188 & 2158056 & 42300840 & 39600010 & 74295022 \\ \hline
Interrupt Time & 1803 & 1598 & 8328 & 163202 & 152779 & 286641 \\ \hline
Perf. Counter & 1803 & 1599 & 8327 & 163204 & 152779 & 286644 \\ \hline
Precise UTC & 1803 & 1598 & 8328 & 163206 & 152779 & 286647 \\ \hline
Tick Count & 0 & 0 & 16 & 16 & 0 & 32 \\ \hline
Multimedia & 0 & 0 & 2 & 16 & 15 & 29 \\ \hline
\end{tabular}
\caption{Dynamic SMC measurements under the same environment and conditions as Table \ref{tbl:staticresult}.}
\label{tbl:dynresult}
\end{table}

We can also inspect SMC-induced pipeline clears directly.  Table
\ref{tbl:smcevents} reports the aggregate counts across $10{,}000$ runs.  The C
variants are affected by context switching and unrelated system activity, which
can include just-in-time execution from other processes.  The naive SMC variant,
unsurprisingly, triggers a large number of clears because the checksum enters its
own introspective region and continues modifying code there.  The unrolled static
and dynamic loop variants remain the most instructive result: they still incur
some clears at the start of each unrolled iteration, showing that the current
termination logic leaves room for improvement, but the reduction relative to the
naive SMC case is dramatic.

\begin{table}[t]
\centering
\small
\begin{tabular}{|c|c|}
\hline
\textbf{Variant} & \textbf{Event count} \\ \hline
C & 4514 \\ \hline
SMC & 83198361 \\ \hline
C Loop & 20986 \\ \hline
SMC Loop & 4492871 \\ \hline
Dyn SMC Loop & 4597855 \\ \hline
\end{tabular}
\caption{SMC pipeline-clear occurrences over $10{,}000$ runs.}
\label{tbl:smcevents}
\end{table}

\section{Efficiency analysis}\label{sec:efficiency}

Any non-SMC reproduction of the mechanism must generally spend more space-time
resource than its SMC counterpart.  Even when both are asymptotically linear, the
relevant practical quantity is the constant factor: the goal is to engineer a
large enough constant $c$ such that
$c_{\text{non-SMC}}\,n \gg c_{\text{SMC}}\,n$.  A
stronger checksum family might even force super-linear behavior in a faithful
non-SMC emulation, though we do not rely on that here.

Precisely proving optimality is difficult because the cost is dominated by
microarchitectural details.  Searching for an optimal non-SMC emulation strategy
is, in general, combinatorial.  At the same time, handwritten assembly is not
automatically optimal: compilers can sometimes discover improvements that are
hard to find manually.  This makes high-level prototyping for code generation a
valid engineering strategy.

Several architectural lessons are important.  First, modifying instructions that
are effectively in flight can reduce performance to something reminiscent of a
late-1990s Pentium~III-class core, because pipeline-clearing behavior destroys the
benefits of deep speculation and out-of-order execution.  Second, branching is
expensive even with a branch predictor, so non-SMC emulations that require many
conditionals are intrinsically disadvantaged.  Third, an ISA-independent design
rule is that SMC should conservatively avoid modifying the currently executing
page.  Finally, timing scores are stable enough in practice for detection, with
the largest perturbations caused by operating-system task switches; those can be
mitigated, though not eliminated, by adjusting process or thread priority.

\section{Using primitives as part of a tamper-proofing scheme}\label{sec:using-primitives}

The model requires a timer $T(t)$ tailored for reliability at the $t$th query and
a checksum function $C(c)$ over code or data $c$.  The tamper-proofing predicate
$P(T,C)$ returns true for the unmodified protected code (when its timing lies within the calibrated bounds) and false for every modification that changes the checksum state or violates the time bound.
In practice, security comes from deploying many checksum units so they cannot all
be removed or overridden easily.  A state machine with precomputed validation
states can implement $P(T,C)$ for a controlled subset of states, while a reset
function restores the SMC to a canonical state and bounds the state-space growth.

The timer-validity thresholds can be estimated empirically on representative
hardware or upper-bounded conservatively.  The most useful policy parameters are
the allowed CPU-usage range in the operating-system environment, the length of
the protected computation, and the number of consecutive failed timing scores that
trigger fatal termination or recovery.  In a practical deployment, these
thresholds are tuned together with code layout rather than in isolation.

\section{Security assumptions and threat model}\label{sec:assumptions}

We state the adversary, its goal, the assumptions the scheme needs, and what is proved.  The statements are formalised and checked in Lean (Section~\ref{sec:formal}); platform behaviour enters only as explicit contract fields (timing bounds, which capabilities a privilege level has), not as proved facts about hardware.

\paragraph{Adversary levels.}
L0: patches the protected bytes before execution.
L1: user mode; may write to the protected process while it runs, single-step or breakpoint it, run it under emulation, and hook operating-system timer interfaces.
L2: kernel or hypervisor; additionally may give instruction fetch and data reads different views of memory and may trap or offset the time-stamp counter.
L4: additionally has physical writers such as DMA devices.
(L3 is reserved and has the capabilities of L2 in the model.)
The \emph{goal} is that a protected byte other than a legitimately varying SMC opcode byte is executed in modified form and no alarm is effective at any point of the (unbounded) run.
An alarm is effective when a checksum unit with at least one response byte, all unmodified, sees a wrong result, or when $k\ge 1$ consecutive checks of one unit exceed its time threshold $\tau$.
The SMC opcode bytes are reset by every check, so tampering with them is outside the goal.

\paragraph{Assumptions.}
A1 (coherence, replaces ``von Neumann architecture''): the fetch and data views are one image.
A2 (coverage, replaces ``cannot neutralize all functions''): every protected byte lies in the region of some checksum unit, each unit's own code is protected, the coverage graph of the units (a unit covers another when its region contains the other's code) is strongly connected, and at least one unit is never modified.
A3 (timing gap, replaces the non-existence of an equally fast non-SMC implementation): the native check time is at most $\tau$, and $\tau$ is below the lower bounds on emulation, on a debug trap, and on mirroring an opcode write.
A4 (no physical writer): no device other than the protected thread writes the protected memory; a larger-area defense narrows the window between check and use but cannot close it.
A5 (time integrity): every timer reading equals the elapsed time.
A6 (persistence): $k\ge 1$, and a modification persists through the next $k$ scheduled checks of each unit that covers it.
A7 (response coverage): each unit has at least one response byte, and its response bytes are part of its own protected code and are not SMC opcode bytes.
A8 (non-compensation): a modification inside the region of a running unit changes the unit's result.
The checksum is deliberately not collision resistant, so A8 is an assumption about the design, and it cannot be dropped: two modified quadwords can restore the final sum and opcode state.

\paragraph{Results.}
(T1) Under A1--A8 and the platform contract, no adversary achieves the goal: if a protected non-opcode byte is ever executed in modified form, an effective alarm occurs somewhere in the run, not necessarily before that execution.
(T2) With separate fetch and data views, a run that follows the SMC kernel's opcode trajectory needs one mirroring event per opcode-changing iteration (except the last), and under A3 and A5 a mediated check admitted by the cost contract that contains such an iteration is a timing failure.  This does not say that every split-view attack is detected.
(T3) Each of the eleven items A1, A2 (coverage, unit code protected, strong connectivity, one unmodified unit), A3, A4, A5, A6, A7 and A8 is necessary within the model: for each there is a well-formed system and an adversary that satisfies the other ten and wins.
(T4) With the inline time-stamp counter, every well-formed timeline at level L0 or L1 satisfies A1, A4 and A5.  Conversely there is a winning adversary against a system that satisfies every assumption except one: A5 at L1 with an operating-system timer interface and at L2, A1 at L2, A4 at L4.  The level capabilities are fields of the contract, not facts about real systems.
With a time source outside the adversary's control, of resolution $r$ and jitter $j$, slowdowns of at least $j+2r$ are detected by a threshold calibrated on native runs.

\paragraph{Scope.}
Not claimed: protection against code injected outside the checksummed bytes, against register tampering outside timed intervals, or against a modification that is undone before the next covering check; guarantees against L2 and above without an external clock; equivalence to cryptographic attestation.
The mechanism is a local runtime-integrity technique intended to raise attacker cost on commodity systems, and the proofs cover the model above, not the hardware.

\section{Summary and outlook}\label{sec:summary}

We have developed a tamper-proofing toolkit that binds code integrity to timing
and runtime-generated SMC.  The central idea is to make faithful reproduction of
both semantics and execution timing costly for an adversary, even when the host
environment is untrusted.

The main engineering lesson is that naive SMC is too slow, but carefully designed
SMC is not.  Reliable clocks such as \texttt{RDTSCP} provide stable enough timing
signals for practical thresholding, while loop unrolling and cross-page
self-modification largely eliminate the worst pipeline stalls.  Dynamic SMC then
turns these gains into a flexible deployment strategy that is materially faster
than a semantics-faithful non-SMC emulation.

The approach is still bounded by platform-specific threshold tuning, operating-
system scheduling noise, and the practical difficulty of measuring some hardware
events without privileged support.  Promising directions for future work include
cross-modifying code in multithreaded settings, checksum kernels whose best
non-SMC emulation incurs super-linear overhead, automated blueprint generation for
dynamic SMC, and validation on additional ISAs.

Overall, polymorphic SMC paired with carefully engineered timing semantics offers
a practical path to runtime integrity for trusted code executing in untrusted
environments.

\section{Formal verification}\label{sec:formal}

\begin{sloppypar}
The threat model of Section~\ref{sec:assumptions} is formalised and machine-checked in Lean~4 (toolchain \texttt{leanprover/lean4:v4.35.0-rc3}, no Mathlib)~\cite{smctamper-lean}.
The Lean development is registered with the Palomar registry as entry \texttt{PALOMAR-2026-10-18-000004}, version~1, and is published at \url{https://github.com/GregoryMorse/SelfModTamper}, release tag \texttt{palomar-v1}, commit \path{0a48a85e0f1884a50c6abdcf8617bd315a6b59c2}.
The registered statement consists of the four results T1--T4 of Section~\ref{sec:assumptions}, proved for the abstract model defined there (checksum units, adversary levels L0--L4, assumptions A1--A8, and a cost contract).
Platform behaviour, that is, what a privilege level can do, which timing bounds hold, and what an operating system or processor provides, enters only as explicit fields of that model, not as a proved fact about real hardware or software.
The same repository contains Lean checks of further statements of this paper (layout and timing arithmetic, checked as consistency of the reported numbers; the separation of Harvard and von Neumann architectures; a forgery of the checksum from three modified quadwords) that are not part of the registered statement.
The repository file \texttt{lean/PAPER\_MAP.md} assigns each claim of the paper to a Lean declaration, a cited platform fact, or the class of observation.
The Lean development is supporting material: the paper's measurements and listings do not depend on it, and what it checks is limited to the statements described above.
The listings, native tests and measurements are evidence for the implementation, not proofs.
The corrections to the text of the first version that this version makes were found while formalising the statements.
\end{sloppypar}
\section*{AI assistance disclosure}

Anthropic Claude Opus 5.5 in high thinking mode and Anthropic Claude Sonnet 5.5 in medium thinking mode, used through the Claude Code command-line agent, assisted with the Lean formalization (statement drafting, proof development and proof repair), the validation scripts, the repository documentation, the comparison of the first version with the formalization, and the drafting of the corrections made in this version.
The Lean kernel checks the proofs; the author reviewed the formal statements, the corrections to the text and this manuscript, and assumes full responsibility for the final work.

\bibliographystyle{unsrt}
\bibliography{biblio}

\end{document}